\documentclass[twocolumn,reprint,prb,aps,amsmath,showpacs,floatfix,superscriptaddress]{revtex4-2}

\usepackage{graphicx}% Include figure files
\usepackage{dcolumn}% Align table columns on the decimal point
\usepackage{mathtools}% include equation features
\usepackage{color}
\usepackage{braket}
\usepackage{hyperref}
\graphicspath{{figures/}}
\usepackage{soul}

\usepackage{algorithmicx}

\usepackage{xr}
\newcommand{\Ham}{H}
\newcommand{\fdag}{f^\dagger}
\renewcommand{\ddag}{d^\dagger}
\newcommand{\Hc}{\mathrm{H.c.}}
\newcommand{\eps}{\varepsilon}

\newcommand{\n}{\hat{n}}
\newcommand{\ntot}{N_{tot}}
\newcommand{\ua}{\uparrow}
\newcommand{\da}{\downarrow}
\newcommand{\ulr}{U_{L/R}}
\newcommand{\der}{\mathrm{d}}

\newcommand{\qubit}{DIQ}

\usepackage[normalem]{ulem}

\newcommand{\qutech}{QuTech and Kavli Institute of Nanoscience, Delft University of Technology, 2600 GA Delft, The Netherlands}

\begin{document}
\title{Two Anderson impurities coupled through a superconducting island: charge stability diagrams and double impurity qubit}

\author{Filip~K.~Malinowski}
\thanks{Present affiliation: Netherlands Organisation for Applied Scientific Research (TNO), P.O. Box 155, 2600 AD Delft, The Netherlands}
\email{filip.malinowski@tno.nl}
\affiliation{\qutech}

\date{\today}

\begin{abstract}
We present a model of two Anderson impurities coupled to and through a superconducting island. The model parametrizes the strength of the coupling between impurity sites, allowing it to represent a variable distance between the impurities. We systematically explore the effect of the model parameters in the subspaces with total even and odd occupancy, identifying unique features of the charge stability diagrams that distinguish the two parities.

For total odd electron parity, we identify a device tuning, in which the splitting between the two lowest energy states is highly insensitive to changes of the chemical potentials. We investigate the degree to which a qubit based on such two states is tunable and discuss an optimal choice of parameters to maximize inhomogeneous dephasing time. Finally, we point out that the proposed qubit lacks a transition dipole moment, and outline some of the consequences on viable driving and readout mechanisms. The prototype qubits could be realized with the existing capabilities of coupling quantum dots to hard-gapped superconductors.

\end{abstract}

\maketitle

\section{Introduction}

The prerequisite to designing a well-performing qubit was to make it highly insensitive to environmental charge noise. Superconducting qubit designs, such as transmons~\cite{koch2007} and fluxonium~\cite{manucharyan2009}, achieve this by using, respectively, large capacitance and large inductance to shunt two superconducting islands. Spin qubits achieve charge-noise insensitivity by relying on a degree of freedom that is (at least in principle) independent of charge~\cite{loss1998}. Last but not least, proposals such as $0-\pi$~\cite{brooks2013} and Majorana qubits~\cite{kitaev2003}, aim to use topological protection to pin the qubit states at zero energy.

Hybrid approaches, which combine superconducting and semiconducting materials, have demonstrated the capacity for coherent manipulation ~\cite{larsen2015,hays2021,pita2022}, including two-qubit coupling~\cite{casparis2016}. However, their performance to date is not sufficient to approach the fidelity thresholds required for quantum error correction. Nevertheless, hybrid structures turned out to be a platform suitable for investigating an extraordinary range of physical phenomena, including but not limited to the multiterminal Josephson effect~\cite{strambini2016,pankratova2020}, superconductor-insulator transition~\cite{bottcher2018} and (signatures of) topological superconductivity~\cite{mourik2012,ren2019,vaitiekenas2020,valentini2021,phan2022}.

A commonly recurring component of super-semiconducting structures is a discrete Fermionic state (e.g. a quantum dot), tunnel coupled to a bulk or mesoscopic superconductor~\cite{pillet2010,lee2014}. Such a component introduces a subgap state into the excitation spectrum, referred to in the literature as Shiba, Yu-Shiba-Rusinov, or Andreev bound state, depending on the device parameters~\cite{estrada2022}. Subgap states can abruptly switch between singlet and doublet ground states (characterized by different total electron parity) under tuning of chemical potential, magnetic field, and so on, constituting a canonical example of a quantum phase transition. The subgap states are studied by their influence on the spectrum of the Josephson junction~\cite{choi2004,cleuziou2006,szombati2016,hays2020,bargerbos2022} and can be combined into chains and molecules~\cite{su2017,grove2018,bouman2020,kurtossy2021,dvir2023}.

In this paper, we describe a system consisting of two Anderson impurities coupled to and through a common superconductor. We enforce the fixed charge on the superconductor and impurities as a whole, which corresponds to the electrostatic floating of the device. Such a configuration prevents conventional transport measurements, and therefore we study the occupancy and quantum capacitance related to the electron tunneling between the impurities and the island. In particular, we identify the differences between the charge stability diagrams with different total electron parities and identify the charge transitions enabling the on-demand generation of entangled spins via Cooper pair splitting.

For total odd electron parity, we identify a device tuning in which the splitting between the two lowest-energy states is highly insensitive to changes of the chemical potential of the impurities. Such insensitivity indicates a potential for the realization of a qubit highly insensitive to charge noise. We investigate the degree to which such a qubit is tunable and discuss an optimal choice of parameters to maximize inhomogeneous dephasing time. Finally, we point out that the proposed qubit lacks a transition dipole moment, and outline some of the consequences on viable driving and readout mechanisms.

We structure the discussion by dividing the article into three main sections. In section~\ref{sec_model} we introduce the model used to simulate the two impurities coupled to a superconducting island. Section~\ref{sec_CSD} systematically analyzes the system: first, removing all coupling terms, and gradually reintroducing them to emphasize effects related to the total parity. Having explored the model, in Sec.~\ref{sec_DIQ} we focus on a specific device configuration, identify the low-energy subspace we propose to use as a qubit, and explore its tunability, possible performance, and some consequences of the vanishing transition dipole moment.

\section{The system Hamiltonian in BCS zero bandwidth approximation}
\label{sec_model}

\begin{figure}[tb]
	\includegraphics[scale=1]{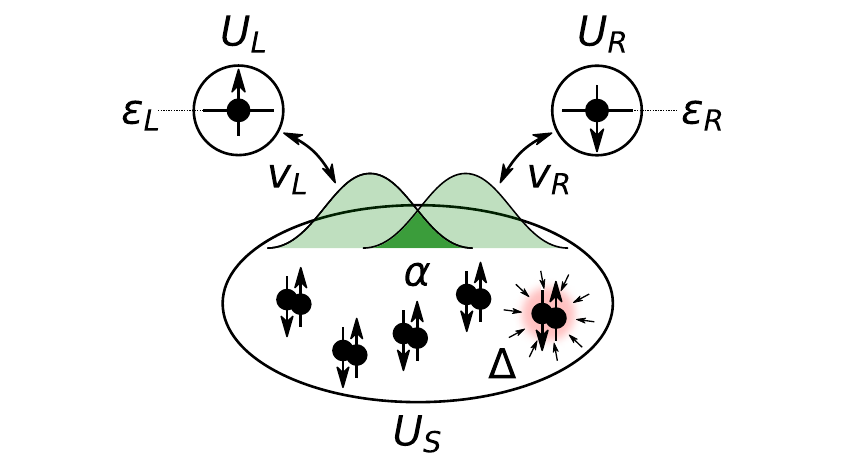}
	\caption{Schematics of the model. Central superconducting island (pairing energy $\Delta$ and charging energy $U_S$) is coupled to two impurity sites (chemical potential $\eps_{L/R}$ and charging energy $\ulr$). The impurity-island coupling is described in terms of tunneling $v_{L/R}$ between impurity sites and superconducting bath orbitals. The strength of coupling between the impurities across the island is parametrized by $\alpha \in [0,1]$ (see main text).
			}
	\label{fig_system_illustration}
\end{figure}

We start by introducing the effective Hamiltonian of Anderson's impurity in the zero-bandwidth BCS approximation, where the superconductor is described as orbitals with a mean-field pairing interaction~\cite{vzitko2022}. The left and right impurity sites ($i = L,R$) correspond to the Fermionic creation operators $\ddag_{i,\sigma}$, with $\sigma = \ua, \da$, and the corresponding electron number operators are denoted as $\n_{i(,\sigma)}$.

The superconductor is represented by two orthogonal orbitals $\fdag_{j,\sigma}$ (where $j=A, B$), each with the pairing term $\Delta \left( \fdag_{j,\ua} \fdag_{j,\da} +\Hc \right)$. The impurity sites are coupled to the superconductor through a term $\fdag_{i,\sigma} d_{i,\sigma}$. We achieve a variable overlap between impurities by choosing $\fdag_{L,\sigma} = \cos(\alpha) \fdag_{A,\sigma} + \sin(\alpha) \fdag_{B,\sigma}$ and $\fdag_{R,\sigma} = \sin(\alpha) \fdag_{A,\sigma} + \cos(\alpha) \fdag_{B,\sigma}$. The transformation is intentionally non-unitary, so that $\alpha \in [0,1]$ represents an effective distance between the impurity sites. The limit $\alpha \rightarrow 0$ represents the distance between the impurity sites much greater than the coherence length of the superconductor, and, conversely, $\alpha \rightarrow 1$ represents the two impurities coupled to the superconductor exactly at the same place.

In summary, we describe the system using a Hamiltonian.
\begin{align}
	\Ham & = \sum\limits_{i=L,R} \left[ \eps_i (\n_i - 1) + U_i (\n_{i} - 1)^2 \right] \nonumber \\
	& + U_S \n_S^2 \nonumber \\
	& + \sum\limits_{i=L,R} \left[ v_i \sum\limits_{\sigma=\ua,\da} \left( \fdag_{i,\sigma} d_{i,\sigma} + \Hc \right) \right] \\
	\label{eq_hamiltonian}
%	& + t \sum\limits_{\sigma=\ua,\da} \left( \ddag_{L,\sigma} d_{R,\sigma} + \Hc \right) \nonumber \\
	& - \Delta \sum\limits_{j=A,B} \left( \fdag_{j,\ua} \fdag_{j,\da} + \Hc \right). \nonumber
.\end{align}
The first row of the Hamiltonian describes the tunable chemical potential $\eps_i$ of the impurities and their charging energy $U_i$. The second row represents the charging energy of the superconducting island $U_S$. Since the total charge is fixed, the charge on the island is expressed in terms of the occupancy of the impurity sites $\n_s = 2 - \n_L - \n_R$.
The third row represents the coupling of the impurity sites to the superconductor $v_i$. The last row represents the superconducting coupling $\Delta$ at the two superconducting orbitals $A, B$.

To illustrate the experimental relevance of the proposal, in the following sections we  employ parameter values relevant for a specific platform -- the semiconducting quantum dots in 2-dimensional electron or hole gas, and the superconducting island is based on an aluminum-proximitized semiconductor. Such a choice sets the Hamiltonian parameters to the typical values of about:
\begin{itemize}
	\item $U_i$ from 0.05 to 5 meV;
	\item $U_S$ from 0 to 5 meV;
	\item $\Delta$ from 0.2 to 0.3 meV.
\end{itemize}
In the remainder of the paper, we set $\Delta=0.25$ meV and vary other parameters. 	

\section{Charge stability diagrams}
\label{sec_CSD}

\subsection{Stability diagrams in absence of tunnel coupling}

\begin{figure}[tb]
	\includegraphics[scale=1]{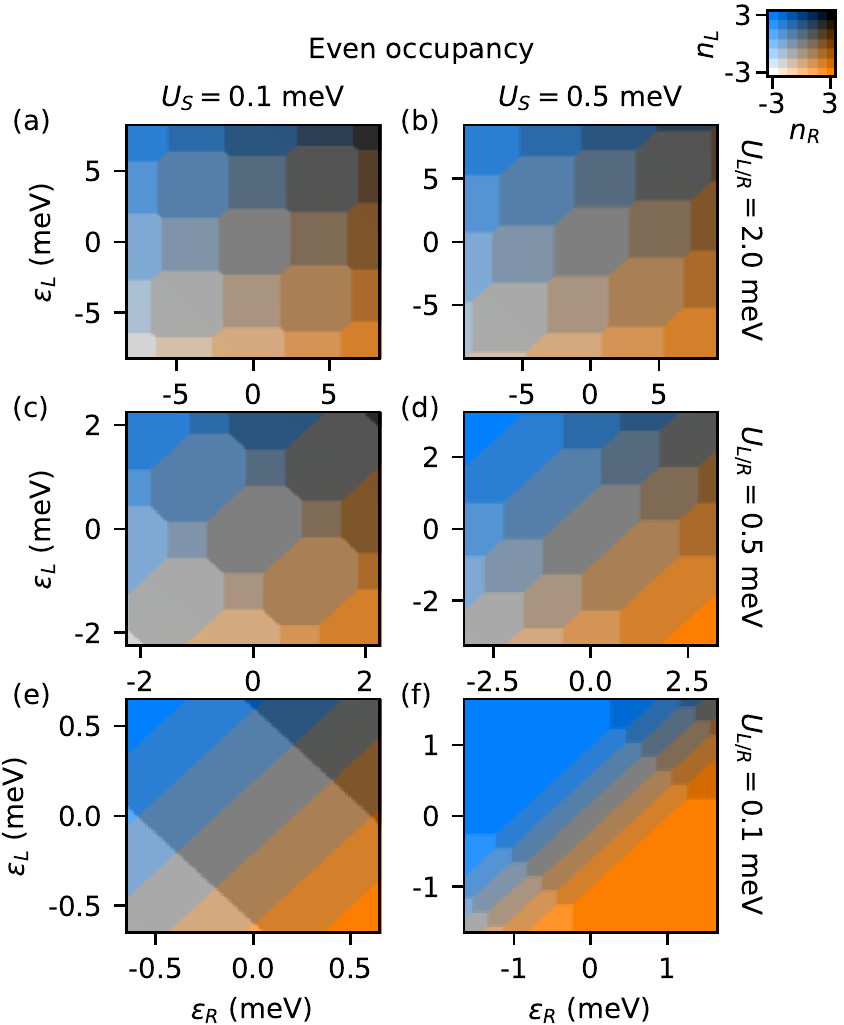}
	\caption{Charge stability diagrams of the floating dot-island-dot, with a total charge fixed ($\ntot=0$). Columns illustrate changes of $U_S$, and rows illustrate simultaneous changes of $\ulr$. A panel in the top right corner presents a color coding of the charge stability regions. In panel (e) all stability regions have an even island occupancy. In panel (f), in the top left and bottom right, the charge is fixed, due to restriction in the occupancy of the dots to at most $\pm 3$ electrons.
			}
	\label{fig_no_tunneling}
\end{figure}

We begin by investigating the interplay between the superconducting gap $\Delta$ and the charging energies $U_{L, R, S}$ while setting $v_i=0$ and $\alpha=0$. Fig.~\ref{fig_no_tunneling}) presents the resulting charge stability diagrams, and the ground state charge dependence on the chemical potential of the left and the right dot $\eps_{L/R}$.

For $\ulr > \Delta > 2\times U_S$ (Fig.~\ref{fig_no_tunneling}(a)) the charge stability regions form a near-square lattice. Vertical (horizontal) charge transitions correspond to moving a single electron between a superconducting island and a right (left) dot. The charge stability regions with an even number of electrons on the island are slightly increased in size, due to the energy reduced by $2\Delta$.

As the value of $\ulr$ decreases, the term $2\Delta$ becomes increasingly more relevant, leading to the shrinking and eventual disappearance of the stability regions with an odd-occupied island (Fig.~\ref{fig_no_tunneling}(c,e)). On the other hand, increasing $U_S$ has the effect of stretching the charge stability regions along the $\eps_L=\eps_R$ axis (Fig.~\ref{fig_no_tunneling}(b,d,f)).

In the absence of tunnel coupling terms, neglecting the level spacing in the dots results in identical charge stability diagrams for total even and odd total occupancies. The only difference is that the centers of charge stability regions with even- and odd-occupied island exchange positions. The exchange of positions, as well as a set of charge stability diagrams for a larger set of $\ulr$ and $U_S$ values, is presented in Supplementary FIg.~\ref{fig_no_tunneling_overview}. Now, we fix $\ulr = 0.5$~meV and $U_S = 0.1$~meV. For these values, both the even and odd charge stability regions are sizable and easily distinguishable, allowing us to illustrate the consequence of nonzero tunnel couplings $v_{L/R}$ and $t$, as well as $\alpha$.

\subsection{Impurity-island coupling in total even and odd occupancy}

Non-zero tunnel couplings predominantly affect the vicinity of the charge transitions, which motivates us to change the focus to the quantum capacitance of the left and the right impurity with respect to the gate voltage that controls its chemical potential $\eps_i = e \zeta_i V_i$. Including the thermal distribution between the eigenstates of the Hamiltonian, the quantum capacitance is given by~\cite{esterli2019,malinowski2022}
\begin{equation}
	C_{q,i} = \frac{1}{Z} \sum\limits_{j} e^{-E_j/\beta} \frac{\partial^2 E_j}{\partial V_{i}^2},
\end{equation}
where $Z = \sum_{j} e^{-E_j/\beta}$ is the canonical partition function, $E_j$ is the energy of the $j$-th lowest state, $\beta=k_B T=10$~$\mu$eV represents the temperature ($T \approx 120$~mK), and the lever arms are set to $\zeta_i=1$.

\begin{figure}[tb]
	\includegraphics[scale=1]{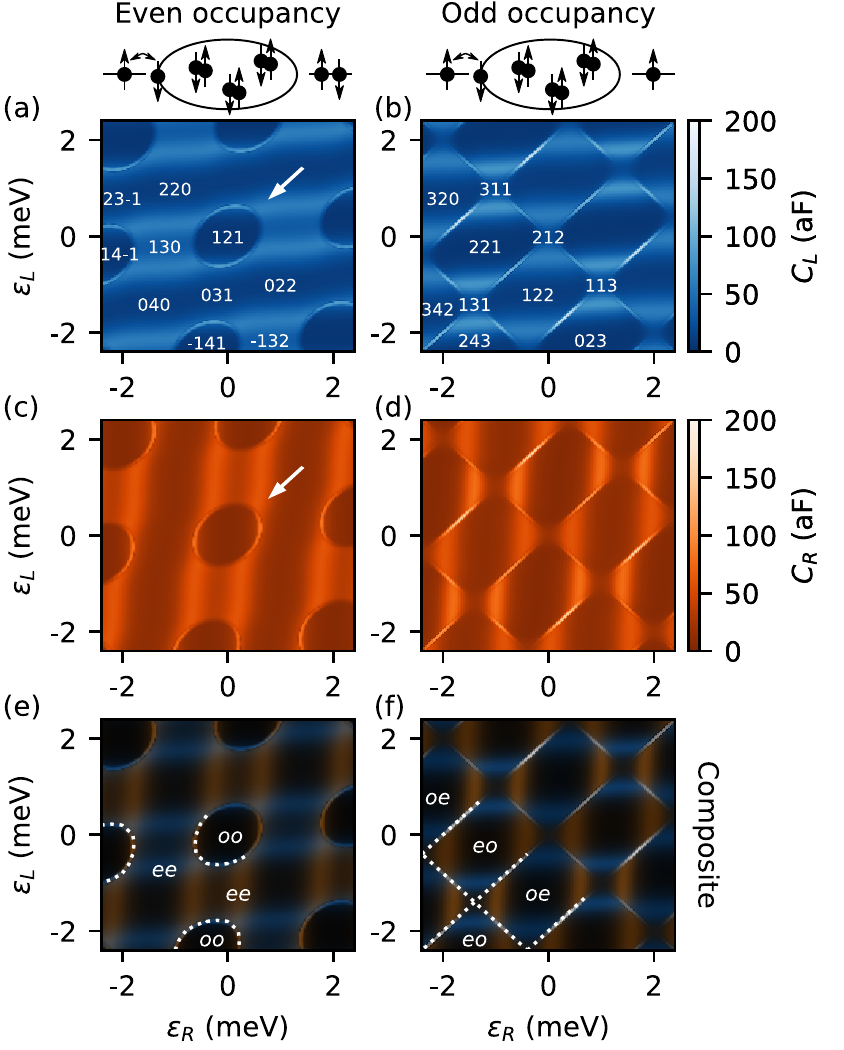}
	\caption{Quantum capacitance on of the dot-island-dot with respect to dot plunger gate voltages in total even (a,c,e) and odd (b,d,f) occupancy. (a,b) and  (c,d) Quantum capacitance of the left and right impurity $ C_{L, R}$. (e,f) Composite of $C_L$ and $C_R$, in blue and orange, respectively. The artifacts near the sharp features in (a,e,c) as well as saturation of the color scale in (b,d,f) are due to numerical differentiation in the vicinity of the crossing between the eigenstates.
			}
	\label{fig_symmetric_tunneling}
\end{figure}

We set a moderately high value of the impurity-island coupling $v_i=0.2$~meV and keep zero overlap between the impurity wavefunctions $\alpha=0$. The resulting charge stability diagrams (Fig.~\ref{fig_symmetric_tunneling}) exhibit a striking difference depending on total parity. In the total even occupancy some of the transitions are broadened, nearly merging the two neighboring charge states. Other charge states (indicated by arrows in Fig.~\ref{fig_symmetric_tunneling}(a,c), and outlined with dotted lines in (e)) remain sharply separated, rounded in shape, and shrink in their extent. 

Differently, in the total odd occupancy, all of the transitions that exchange an electron between one of the dots and the island are broadened, and none of the stability region shapes is rounded. The diagonal two-electron transition (with positive and negative slopes) is sharp and divides the gate voltage space into rectangles (as outlined with dotted lines in Fig.~\ref{fig_symmetric_tunneling}(f)).
Supplementary Fig.~\ref{fig_symmetric_tunneling_overview} illustrates that the differences between the total oven and odd occupancy are not a consequence of fine-tuning, but develop gradually as the tunnel coupling $v_i$ increases.

The sharp charge transitions indicated in Fig.~\ref{fig_symmetric_tunneling}(f) originate from the separation of the Hilbert space into two subspaces, differing in the parity of the impurity sites. The ground state always consists of a doublet localized on one impurity and a singlet on another ($eo$ and $oe$), but these subspaces do not couple due to the lack of crossed Andreev reflection and cotunneling through the island ($\alpha=0$).

For even total occupancy (Fig.~\ref{fig_symmetric_tunneling}(e)), the rounded region ($oo$) corresponds to the 4-degenerate ground state with both impurities in a doublet state. In the surrounding area ($ee$) both impurities are in a singlet state, with all electrons paired on an island, impurity, or on a formed subgap states.

\subsection{Overlap of impurity wavefunctions}

\begin{figure}[tb]
	\includegraphics[scale=1]{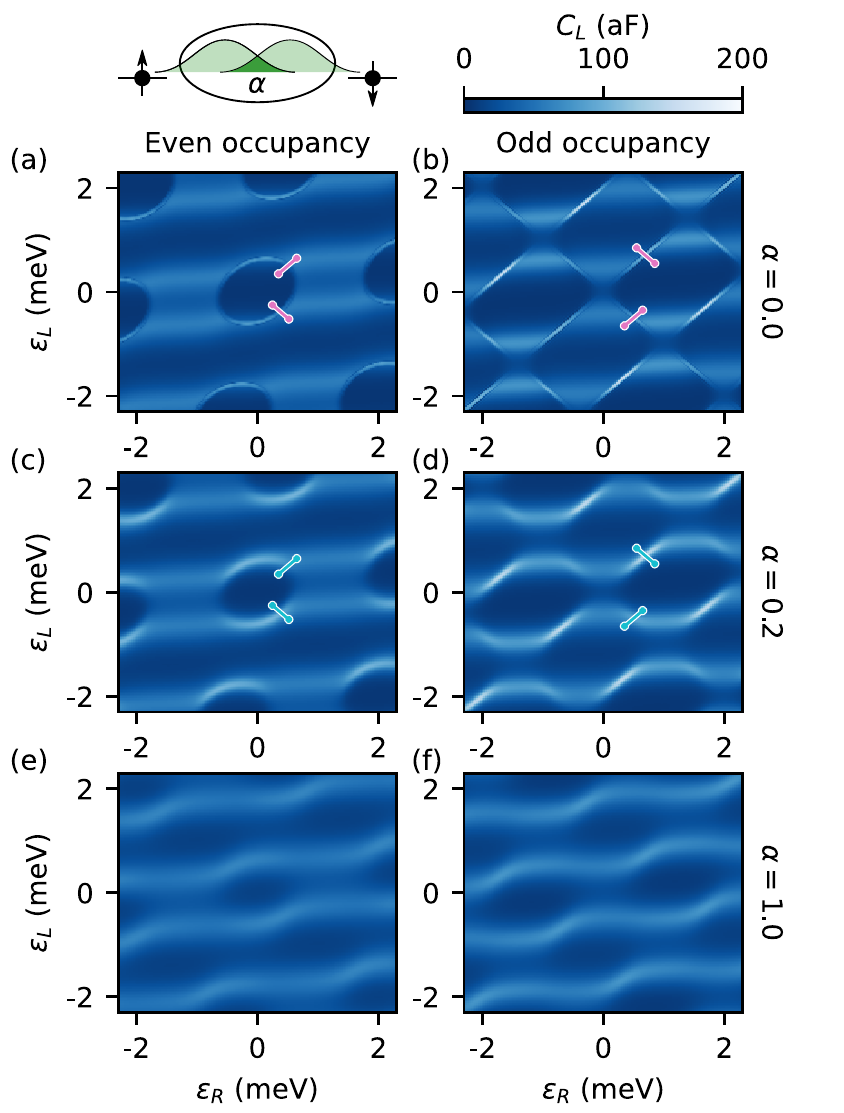}
	\caption{Quantum capacitance with respect to left gate voltage for (a,b) $\alpha = 0$, (c,d) $\alpha = 0.2$ and (e,f) $\alpha = 1$. Colored line segments correspond to energy diagrams illustrated in Fig.~\ref{fig_opening_tunneling}.
			}
	\label{fig_alpha_tunneling}
\end{figure}

To complete the exploration of the charge stability diagrams, we add an overlap between the electronic wavefunctions localized at the two impurity sites. Increasing the value of $\alpha$ from 0 to 0.2 and 1 (Fig.~\ref{fig_alpha_tunneling}) allows tunneling of the quasiparticle between the two impurity sites. In the total even occupancy, this introduces coupling between $ee$ and $oo$ subspaces, and in the total odd occupancy, between $eo$ and $oe$ subspaces (c.f. Fig.~\ref{fig_symmetric_tunneling}(e,f)). Increasing $\alpha$ from 0 to 0.2 creates an avoided crossing between those subspaces, prevents thermal excitations, and allows the detection of the quantum capacitance of the ground state~\cite{esterli2019}. As a result, the corresponding charge transitions broaden, and the measured capacitance increases in magnitude. A further increase in $\alpha$ broadens the charge transitions even more but reduces the capacitance magnitude. 

\begin{figure}[tb]
	\includegraphics[scale=1]{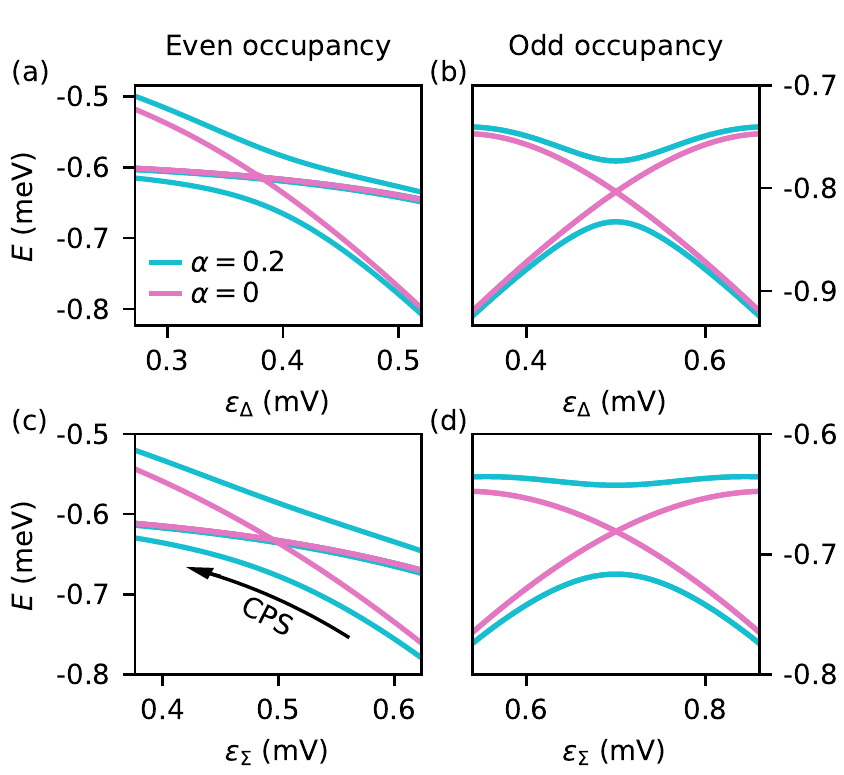}
	\caption{Energy diagrams in the vicinity of the charge transitions between (a,c) $oo$ and $ee$ ground states, and (b,d) $eo$ and $oe$ ground states. Panels (a,b) correspond to the elastic co-tunneling process, while (c,d) crossed Andreev reflection.
			}
	\label{fig_opening_tunneling}
\end{figure}

In the charge stability diagrams, we highlight transitions of interest, indicated with colored line segments. The corresponding energy diagrams are shown in Fig.~\ref{fig_opening_tunneling}. The transitions indicated by the negative diagonal line segments in Fig.~\ref{fig_alpha_tunneling} correspond to elastic cotunneling of a single quasiparticle between left and right impurities (Fig.~\ref{fig_opening_tunneling}(a,b)). In these cases, the energy diagram is analogous to that of a normal double quantum dot with direct tunnel coupling~\cite{petta2005}, or coupling via an intermediate quantum state~\cite{baart2017}.

For total even occupancy, the ground state is a non-degenerate spin singlet, and the lowest excited state is a triply degenerate spin triplet. The splitting between the singlet and triplet subspaces is minimized in the $oo$ charge stability region, where the wavefunctions of the two odd occupied impurities overlap the least and are maximized in the $ee$ charge stability region, where the spin tripled state is in Pauli spin blockade, analogous to a conventional double quantum dot. In contrast, in total odd occupancy, the doubly spin-degenerate undergo avoided crossing at between $eo$ and $oe$ charge stability regions without spin blockade taking place.

The transitions along positive diagonal line segments in Fig.~\ref{fig_alpha_tunneling}(a-d) correspond to the splitting of the Cooper pair into two quasiparticles. The corresponding energy diagrams (Fig.~\ref{fig_opening_tunneling}(c,d)) have an identical structure to the co-tunneling transitions (Fig.~\ref{fig_opening_tunneling}(a,b). The identical structure of the energy diagrams suggests a concrete realization of the on-demand cooper splitting by the adiabatic crossing of the charge transition with a baseband pulse tuning the chemical potential of the two impurities simultaneously. Such a protocol requires only the means of tuning the chemical potential, such as those in Ref.~\cite{petta2005}. The protocol was hinted at in Ref.~\cite{dejong2022} and is likely to be possible to execute in setups exhibiting high efficiency of Cooper pair splitting \cite{wang2022wire,wang20222DEG}. However, the direct Bell test on the ejected spin would require implementation in a material that enables high-fidelity single-spin rotations and single-shot readout.

\section{The double-impurity qubit}
\label{sec_DIQ}

For a total-odd occupancy, a point of particular interest lies at the crossing of dotted-white lines. In the remainder of this article, we present that in the absence of subgap state overlap, the degeneracy between the $eo$ and $oe$ states enjoys a high degree of protection from the charge noise. Building on that observation and reintroducing the wavefunction overlap, we propose a qubit realization based on the location of the quasiparticle that can enjoy near third-order protection against charge noise.

\begin{figure}[tb]
	\includegraphics[scale=1]{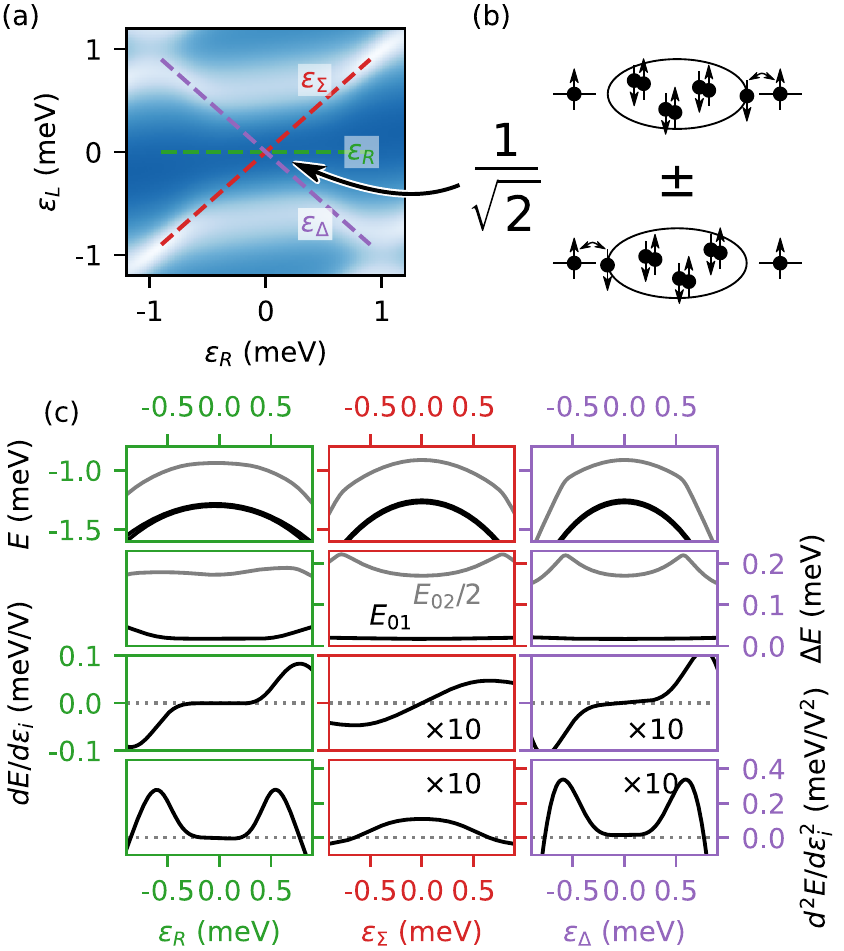}
	\caption{(a) Zoom in at the sweet spot of the \qubit, measured as a capacitance of the left impurity. Dashed lines indicate the axis of the plots in panel (c).
	(b) Cartoon illustration of the qubit eigenstates at the sweet spot.
	(c) Characterization of the qubit spectra at the sweet spot along $\eps_{R,\Sigma,\Delta}$ axes.
		The first raw presents the spectra of the qubit states (black) and the lowest non-computational state.
		The second row illustrates the qubit splitting $E_{01}$ and half the splitting to the lowest non-computational state $E_{02}/2$.
		The third and fourth rows present the first and second derivative of the qubit energy with respect to $\eps_{R,\Sigma,\Delta}$, i.e. $\der E_{01}/ \der \eps_i$ and $\der^2 E_{01}/ \der \eps_i^2$.
			}
	\label{fig_qubit_sweet_spot}
\end{figure}

\subsection{A sweet spot with near-third order protection}
\label{sec_sweet_spot}

We proceed to identify a point in the charge stability diagram for a total odd electron occupancy where the splitting between the two lowest-energy states exhibits exceptionally weak dependence on the chemical potentials. Such a point would mark a potential for implementation of a qubit with a relatively long coherence time. The promising sweet spot lies at the intersection of positive and negative diagonal lines that separate the ground state $eo$ and $oe$ (cf. Fig.~\ref{fig_symmetric_tunneling}(f)). Fig.~\ref{fig_qubit_sweet_spot} presents the charge stability diagram in the vicinity of the sweet spot. We dedicate this section to the analysis of its properties and tunability.

The two lowest eigenstates at the sweet spot are, illustratively, equal superpositions of $\ket{R} = \tfrac{1}{\sqrt{2}}\left( \ket{\sigma, 2, 0} - \ket{\sigma, 0, 2}\right)$ and $\ket{L} = \tfrac{1}{\sqrt{2}}\left( \ket{2, 0, \sigma} - \ket{0, 2, \sigma}\right)$, where the three indices represent the occupancy of the left dot, island, and the right dot (up to $2N$; $\sigma$ indicates a spin of a singly occupied dot). The two states correspond to a single quasiparticle occupying one of the two impurity sites, similar to an electron in the case of a charge qubit, or a Cooper pair in the case of a Cooper pair box qubit. A critical feature that distinguishes the proposed sweet spot is the introduction of the superconducting island. The electron-hole symmetry introduced by the island suppresses the charge dispersion, in the sense that $\langle \hat{n}_{L/R} \rangle = 1$ for $\ket{L}$, $\ket{R}$, and \emph{any superposition} of these states.

The top row in Fig.~\ref{fig_qubit_sweet_spot}(c) presents an energy diagram of the three lowest states, the qubit, and the lowest non-computational state, with respect to $\eps_R$, $\eps_\Sigma=(\eps_L+\eps_R)/2$ and $\eps_\Delta=(\eps_L-\eps_R)/2$ for $\Delta=0.25$~meV, $\ulr = 0.5$~ mV, $U_S=0.1$~meV, $v_i = 0.5$~meV and $\alpha=0.02$. The second row illustrates that the qubit splitting (black) is minimal at the sweet spot ($E_{01} = 17.2$~$\mu$eV; $E_{01}/h = 4.16$~GHz) and over an order of magnitude smaller than the energy of the lowest non-computational state ($E_{02} = 357$~$\mu$eV, $E_{02}/h = 86.3$~GHz). Due to left-right and electron-hole symmetries, the splitting at the sweet spot is first-order insensitive to the chemical potential of the dots individually, as well as to their difference and common mode, as illustrated in the third row by depicting the derivatives of the qubit splitting with respect to $\eps_{R/\Sigma/\Delta}$~\footnote{
The derivatives are to be understood as taken along the axis illustrated in Fig.~\ref{fig_qubit_sweet_spot}, that is, $\left. \partial d^n E_{01}/\partial \eps_i^n \right|_{\eps_i'=0}$ with $\eps_{R'}=\eps_{L}$, $\eps_{\Sigma'}=\eps_{\Delta}$ and $\eps_{\Delta'}=\eps_{\Sigma}$.
}. The notable feature of the qubit is that the second derivative of qubit splitting with respect to gate voltages is also exceptionally small. This is in contrast to charge qubits and Cooper pair box qubits, characterized by very short coherence times, for which the second derivative is maximized at the first-order sweet spot.

We propose that the nearly third-order insensitivity to chemical potential makes $\ket{L}$ and $\ket{R}$ states promising for the realization of a highly coherent qubit, and the charge protection is qualitatively different from that for superconducting, spin, and topological qubits. Next, quantify the possible coherence time of the qubit in the presence of quasistatic noise and investigate the optimal choice of $\ulr$, $\alpha$, and $v_i$.

\subsection{Qubit tunability in microwave frequency range}
\label{sec_tuning}

\begin{figure}[tb]
	\includegraphics[scale=1]{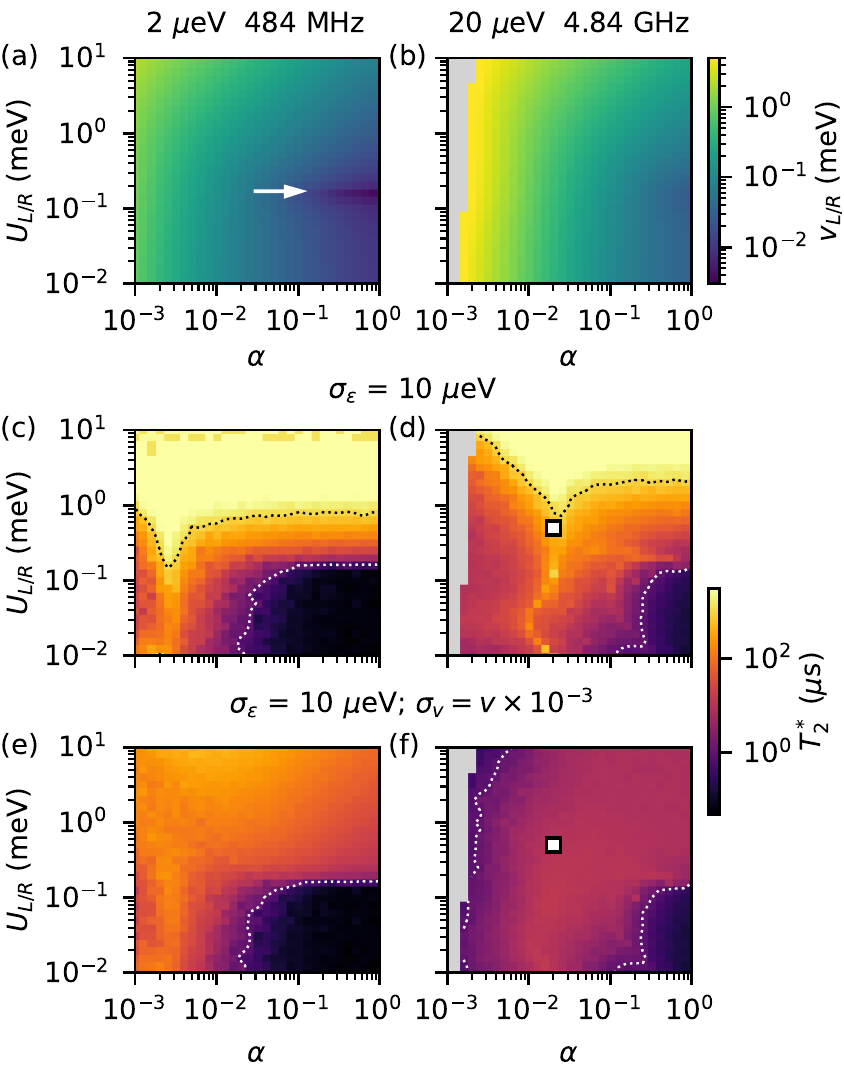}
	\caption{Coupling between the impurities required to achieve the splitting between the qubit states of (a) 2~$\mu$eV and (b) 20~$\mu$eV. (c-f) The pure dephasing time $T_2^*$ for given $\ulr$ and $alpha$ and $v_{L/R}$ adjusted to achieve the splitting of (c,e) 2~$\mu$eV and (d,f) 20~$\mu$eV. Dotted white and black lines mark $T_2^* = 1$~$\mu$s and 1~ms, respectively. In (c,d) Gaussian quasistatic noise affecting independently the chemical potential $\eps_{L/R}$ with magnitude $\sigma_\eps=10$~$\mu$eV. In (e,f) we also include Gaussian quasistatic noise on the dot-island coupling terms with magnitude $\sigma_v = v_{L/R} \times 10^{-3}$. The top three rows in the color map (c) exhibit numerical artifacts due to issues estimating the longest dephasing times
			}
	\label{fig_tunability}
\end{figure}

A realization of the \qubit~would consist of dots or impurities of a fixed charging energy at a fixed distance. Thereby we address the question, for fixed $\ulr$ and $\alpha$, is it always possible to tune the qubit to arbitrary energy (frequency) within the range conveniently accessible by microwave driving?

As an example, we select representative qubit splittings of $E_{01}=2$ and 20~$\mu$eV, corresponding to the frequencies of $f_{01}\approx 484$~MHz and 4.84~GHz, respectively. Figure~\ref{fig_tunability}(a,b) presents the values of $v_{L/R}$ required to tune those qubit frequencies for a range of $\ulr$ and $\alpha$ values and for fixed $\Delta=0.25$~meV, $U_S=0.1$~meV. The values of $\ulr$ and $\alpha$ for which tuning to the desired frequency requires $v_{L/R} > 5$~meV are indicated in gray.

In general, the reduced overlap of superconducting orbitals $\alpha$ requires compensation by increasing $v_{L/R}$. The only qubit configuration that is not accessible with a constraint $v_{L/R}<5$~meV is the large qubit splitting for the smallest values of alpha $\alpha$. We do not expect that to be an obstacle since we demonstrate further that this regime is not optimal for maximizing the inhomogeneous dephasing time.

Otherwise smooth map of the $v_{L/R}$ values has one characteristic feature, for 2~$\mu$eV qubit splitting, indicated by a white arrow in Fig.~\ref{fig_tunability}(a). This feature corresponds to $\ulr=0.175$~meV and indicates the particularly small values of $v_{L/R}$ required to induce qubit splitting.

A particular value of $\ulr=0.175$~meV corresponds to $U_L+U_R = U_D+\Delta$. In the absence of tunnel couplings, such a choice of parameters would lead to the degeneracy between even- and odd-occupied islands, namely the exact values of $U_{L/R/S}$ at which the smaller stability regions, as in Fig.~\ref{fig_no_tunneling}, have shrunk to zero sizes.

\subsection{Pure dephasing due to charge noise}
\label{sec_dephasing}

Now, we illustrate the degree of impact of charge insensitivity on the \qubit~performance. In the analysis, as a figure of merit, we choose an inhomogeneous dephasing time under the influence of quasistatic noise.

First, we consider exclusively quasistatic noise that independently affects the chemical potential $\eps_{L/R}$ of the impurities, characterized by a Gaussian distribution with an RMS magnitude of $\sigma_\eps = 10$~$\mu$eV. The magnitude of the noise is chosen according to the experimental results on gate-defined spin-qubits~\cite{dial2013,martins2016}. Although the assumed noise is Gaussian, it couples to the qubit frequency via a highly nonlinear dependence. Thus, we expect the Ramsey decay to be characterized by an envelope very different from the usual functional form of $\exp [ -(t/T_2^*)^\gamma ]$. Therefore, we sample 500 values of the qubit frequency around the qubit sweet spot. We identify $T_2^*$ as the time scale at which the Ramsey oscillations decay by a factor of $e$.

\begin{figure}[tb]
	\includegraphics[scale=1]{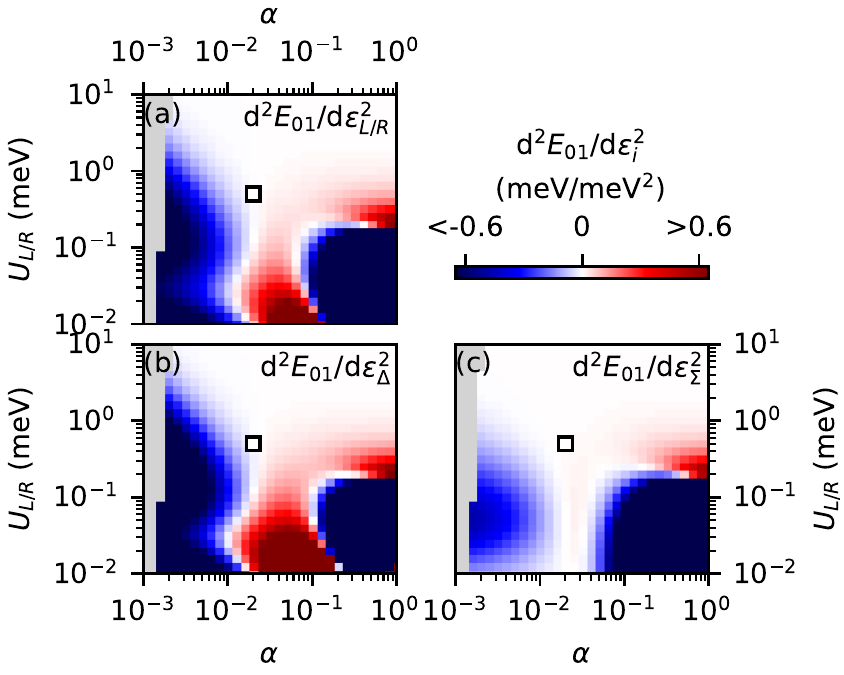}
	\caption{Computed second derivatives of the qubit splitting with respect to a combination of the chemical potential variations. The saturated regions indicate large derivatives and correspond to low predicted dephasing times in Fig.~\ref{fig_tunability}(d).
			}
	\label{fig_d2V}
\end{figure}

Figure~\ref{fig_tunability}(c,d) presents the maps of $T_2^*$ for the qubit tuned to the splitting of 2 and 20~$\mu$eV and suggests the most optimal and most ill-advised parameter choices.
The unfavorable tuning is the regime of small $\ulr$ and large $\alpha$, appearing as a dark region in the bottom right of the color maps. In that regime, the unpaired quasiparticle is strongly localized on the impurity sites. The quasiparticle orbitals on the superconductor have high energy and only mediate the co-tunneling and crossed Andreev reflection through virtual tunneling. As a consequence, qubit splitting and its second derivatives $\der^2 E_{01}/\der \eps_{L/R/\Sigma/\Delta}^2$ have local extrema at the sweet spot. We illustrate this in Fig.~\ref{fig_d2V}, which plots the second derivatives at the sweet spot for $E_{01}=20$~$\mu$eV. The presence of the local extrema is similar to a charge and Cooper pair box qubits and leads to the estimated dephasing times $< 1$~$\mu$s. 

Reducing $\alpha$ while increasing $v_{L/R}$, for small $\ulr<0.1$~$\mu$eV, increases the coherence time. In particular in a range of $\ulr$ values we estimate $T_2^*>100~\mu$s (bright vertical feature at about $\alpha=0.003$ and 0.02 in Fig.~\ref{fig_tunability}(c,d)). Compared to Fig.~\ref{fig_d2V}, we identify the region of increased coherence with reduced second derivatives of the qubit splitting. A square marker points to an approximate choice of the parameters in Fig.~\ref{fig_qubit_sweet_spot}.

Increasing $\ulr$ leads to a further increase in the estimated $T_2^*$, over 1~ms. Accordingly, the second derivatives in Fig.~\ref{fig_d2V} almost vanish. A physical system this regime represents consists of, for example, two tiny impurities (e.g. individual atoms) on a superconductor, forming two deep Yu-Shiba-Rusinov states. In such a case, the screening by the superconductor provides insensitivity to an external electric field.

Alternatively, we consider the noise that affects the chemical potential and the tunnel coupling between the impurity sites and the superconducting sites. As an illustration, we consider the magnitude of tunnel coupling noise proportional to coupling $\sigma_v = v \times 10^{-3}$. This assumption is motivated by studies of gate-defined spin qubits, reporting a nearly exponential dependence of the exchange coupling on gate voltage~\cite{maune2012,dial2013,martins2016}. 

Simulations that include tunnel coupling noise indicate much shorter coherence times, only up to 100 and 10~$\mu$s for $E_{01}=2$ and 20~$\mu$eV, respectively (Fig.~\ref{fig_tunability}). These coherence limits are highly speculative, since they follow directly from the assumption that the noise in tunnel coupling is proportional to the coupling. The simulation however draws an attention to the fact that noise affecting the coupling terms is likely to become the limiting factor once there is sufficient protection against noise in the chemical potential~\cite{martins2016,reed2016}.

\subsection{Vanishing dipole moment, some of the consequences and readout}
\label{sec_dipole}

We now take a step back to investigate the consequences of the electron-hole symmetry introduced by the use of the superconducting island. Together, electron-hole symmetry and tuning to the sweet spot imply that the qubit states are \emph{chargeless}. Specifically, within the qubit subspace (including superpositions of eigenstates), there is an equal probability of finding the impurity site occupied by $1+N$ and $1-N$ electrons. As a result, the expected $\langle \n_{L/R} \rangle = 1$ for all qubit superposition states, implying zero electric dipole and transition dipole moment. 

The first consequence of the vanishing dipole moment is that qubit driving requires a strong magnitude of rf excitation. Only strong excitation would enable a two-photon transition, with a quadratic dependence of Rabi precession on the drive magnitude.

Another consequence is the lack of the qubit-resonator coupling in a Jaynes-Cummings Hamiltonian, the lack of the conventional dispersive shift, and the need for an alternative approach to the qubit readout. A possible readout scheme could combine a baseband pulse, to move the qubit away from the sweet spot, with a readout tone applied to a resonator coupled with one of the impurities. Another scheme could exploit the shift of the resonator due to the residual quantum capacitance difference between the qubit states (Fig.~\ref{fig_d2V}). The third candidate is to employ the quantum capacitance to perform parametric driving of the resonator coupled capacitively to one of the impurities or the island. The latter approach is the least conventional, and therefore we elaborate on how it is to be understood and sketch a protocol it would require.

In brief, the quantum capacitance represents the rearrangement of charge under the application of the gate voltage. At the same time, a change of the charge displaces an electric field in the coupled cavity. This mechanism realizes a readout scheme introduced in Ref.~\cite{didier2015}, and is of interest from the perspective of both -- principle and application. The relation to that proposal is as follows:

For the qubit tuning presented in Fig.~\ref{fig_qubit_sweet_spot}, we identify the second derivative $\der^2 E_{01}/\der \eps_\Sigma^2$ as equivalent to quantum capacitance -- the simultaneous change of the chemical potential at the two impurity sites will move the charge from the impurities to the superconducting island. Let us consider the central island to be capacitively coupled to the resonator. An in-phase drive to the gates that controls the impurity sites, at the frequency of the resonator, would induce a state-dependent excitation in the resonator. Excitation emitted from the resonator to the feedline would provide a signal that measures the qubit state.

\section{Outlook}

To summarize, we have introduced a model of the two impurities coupled to a superconducting island. The model parameterizes the strength of the coupling between the impurity sites, allowing it to represent a variable distance between the impurities. The analysis is limited to zero magnetic fields, zero spin-orbit, fixed parity, and no direct tunneling between the impurities. These constraints can easily be lifted by introducing additional terms to the Hamiltonian, without increasing the size of the Hilbert space. This would extend the applicability of the simulations to Andreev molecules~\cite{su2017,grove2018,bouman2020,kurtossy2021}, including Poor Man's Majoranas~\cite{leijnse2012,dvir2023}, to understand their potential use for prototype qubits.

The double impurity qubit represents a paradigm for achieving protection from charge noise, by taking advantage of the capacity of a superconducting to provide screening while preserving parity of the impurity. The prototype qubits could be realized with the existing capabilities of coupling quantum dots to hard-gapped superconductors. Follow-up theoretical research may consider whether discrete subgap states can provide benefits with regard to susceptibility to noise and investigate more realistic noise models and properties of the qubits in the vicinity of the sweet spot and under small deviations from left-right and electron-hole symmetry. Theoretical investigation or experimental realization may reveal that \qubit~is not a viable alternative platform for quantum computation; nonetheless, the lack of dipole moment makes it a great platform for exploring the effects of multi-photon driving and strongly nonlinear coupling to rf resonators. Last but not least, we note that \qubit~represents the minimal realization of a fermionic quantum bit, as defined by Bravyia and Kitaev~\cite{bravyi2002}.

\section*{Acknowledgments}

I thank Rok~\v{Z}itko for suggestions regarding modeling the double-impurity; Chun-Xiao Liu, Christian Prosko, and Guanzong Wang for discussion of sweet-spots, Poor Man's Majoranas, and longitudinal couplings during the preparation of the manuscript.

This work was supported by NWO under a Veni grant (VI.Veni.202.034).

\bibliography{biblio}

\end{document}

\title{Supplementary simulations to ``Two Anderson impurities coupled through a superconducting island: charge stability diagrams and double impurity qubit''}

\author{Filip~K.~Malinowski}
\email{f.k.malinowski@tudelft.nl}
\affiliation{\qutech}

\date{\today}

\begin{abstract}
\end{abstract}

\maketitle

%\appendix
%\counterwithin{figure}{section}

This appendix includes a superset of simulations presented in the main text.

\begin{figure*}[tb]
	\includegraphics[scale=1]{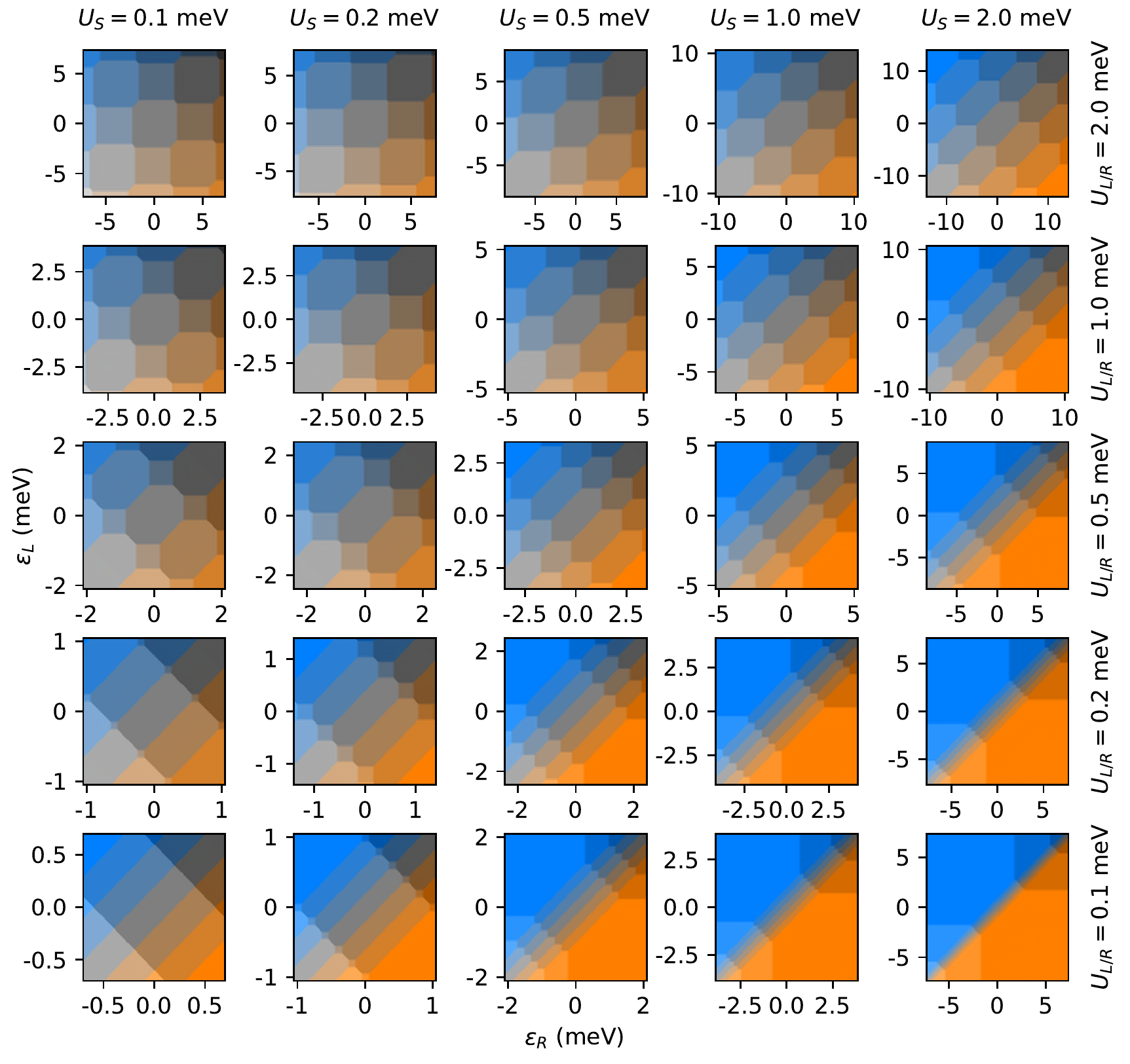}
	\caption{Charge stability diagrams of the floating dot-island-dot, with a total charge fixed ($\ntot=0$). Columns illustrate changes of $U_S$, and rows illustrate simultaneous changes of $\ulr$. A panel in the top right corner presents a color coding of the charge stability regions. The figure illustrates a range of charging energies extended relative to Fig.~2..}
	\label{fig_no_tunneling_overview}
\end{figure*}

\begin{figure*}[tb]
	\includegraphics[scale=1]{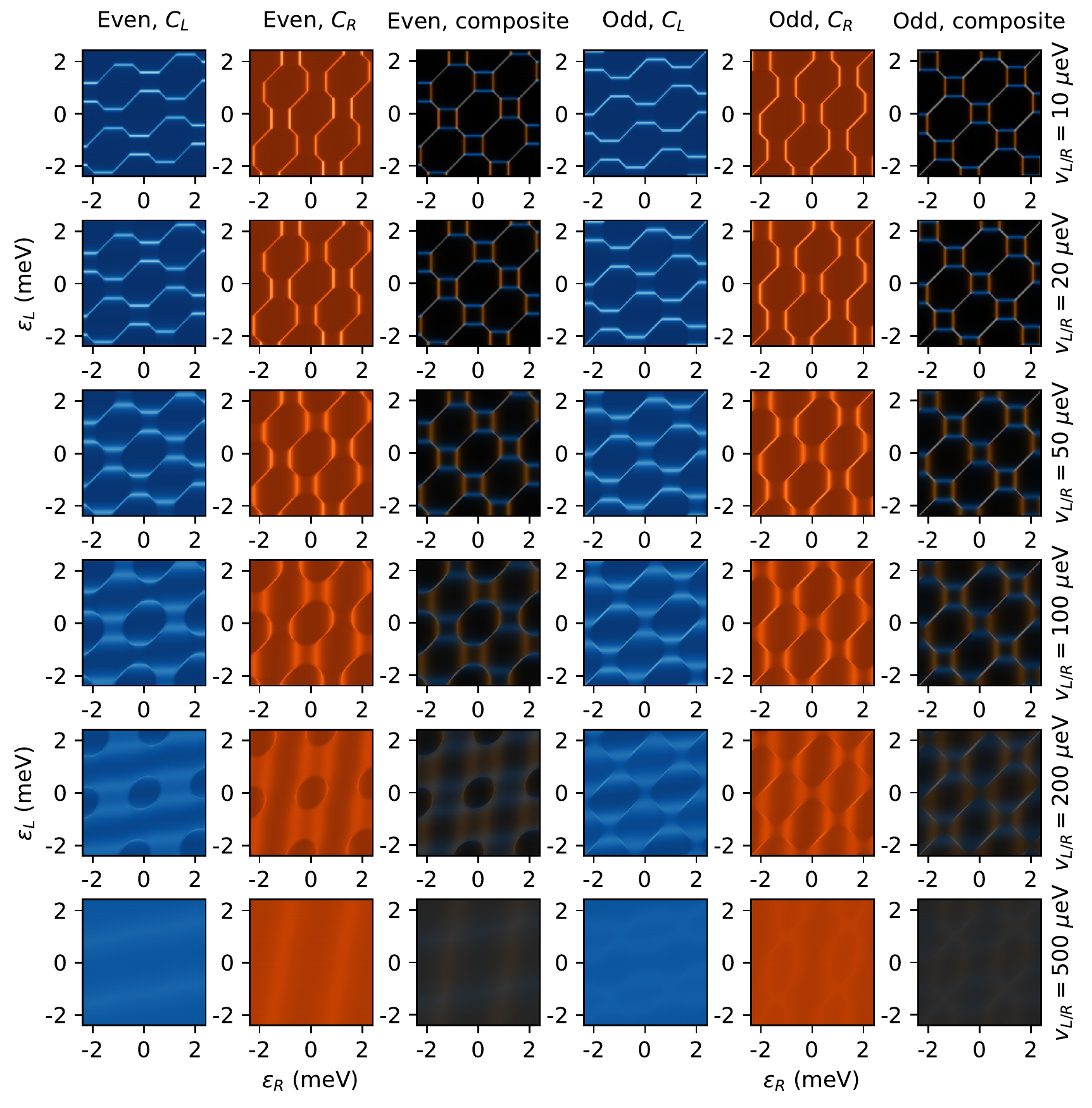}\\
	\caption{Quantum capacitance on of the dot-island-dot with respect to dot plunger gate voltages in total even and odd occupancy, and in a range of symmetrically varied island-impurity tunnel couplings. Blue and orange color maps represent the quantum capacitance of the left and right impurity. The black color maps are composite of the two. Fourth row corresponds to the data presented in Fig.~3}
	\label{fig_symmetric_tunneling_overview}
\end{figure*}

\begin{figure*}[tb]
	\includegraphics[scale=1]{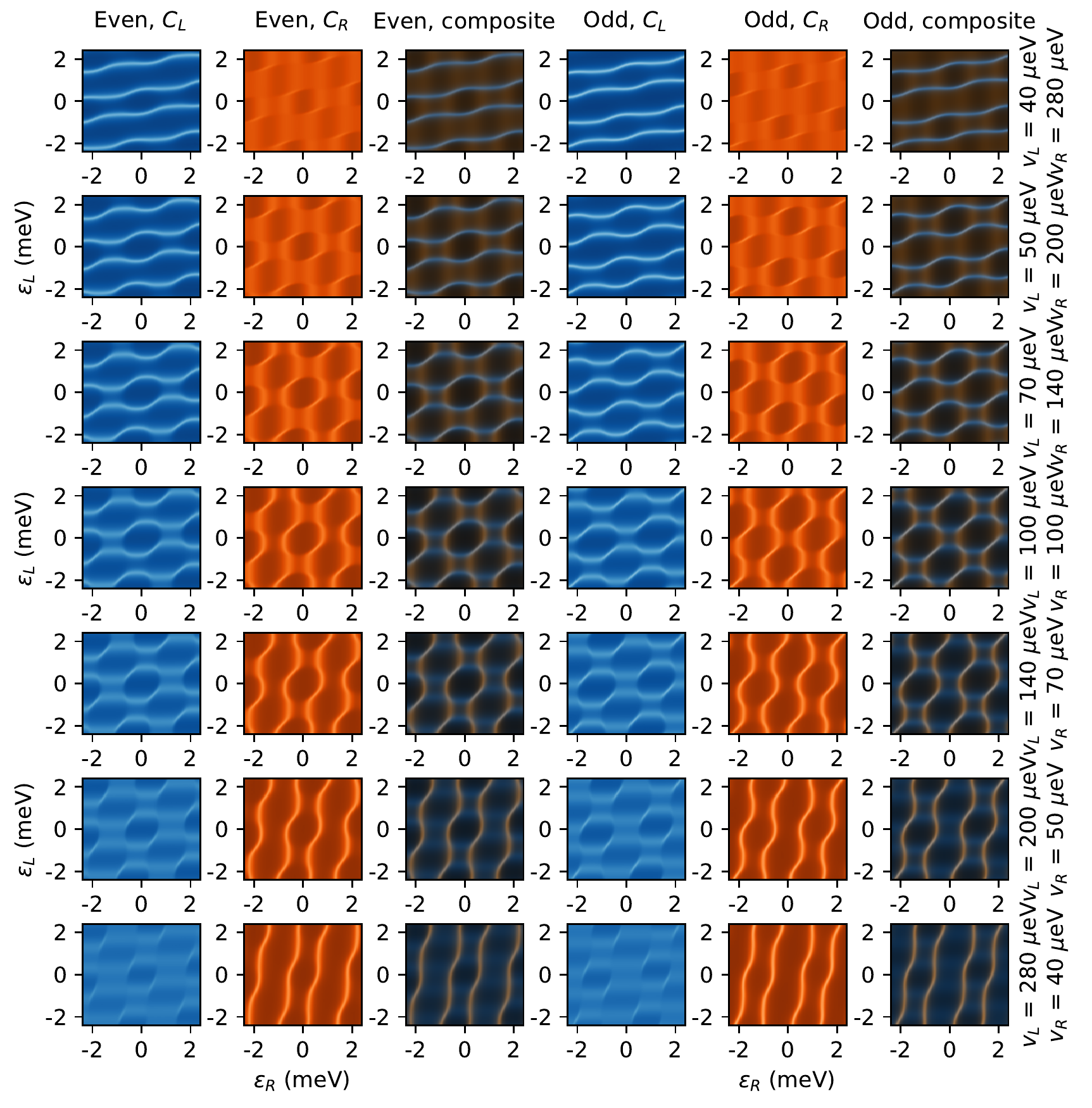}\\
	\caption{Quantum capacitance on of the dot-island-dot with respect to dot plunger gate voltages in total even and odd occupancy, and in a range anti-symmetrically varied island-impurity tunnel couplings. Blue and orange color maps represent the quantum capacitance of the left and right impurity. The black color maps are composite of the two. Fourth row corresponds to the data presented in Fig.~3}
	\label{fig_asymmetric_tunneling_overview}
\end{figure*}

\begin{figure*}[tb]
	\includegraphics[scale=1]{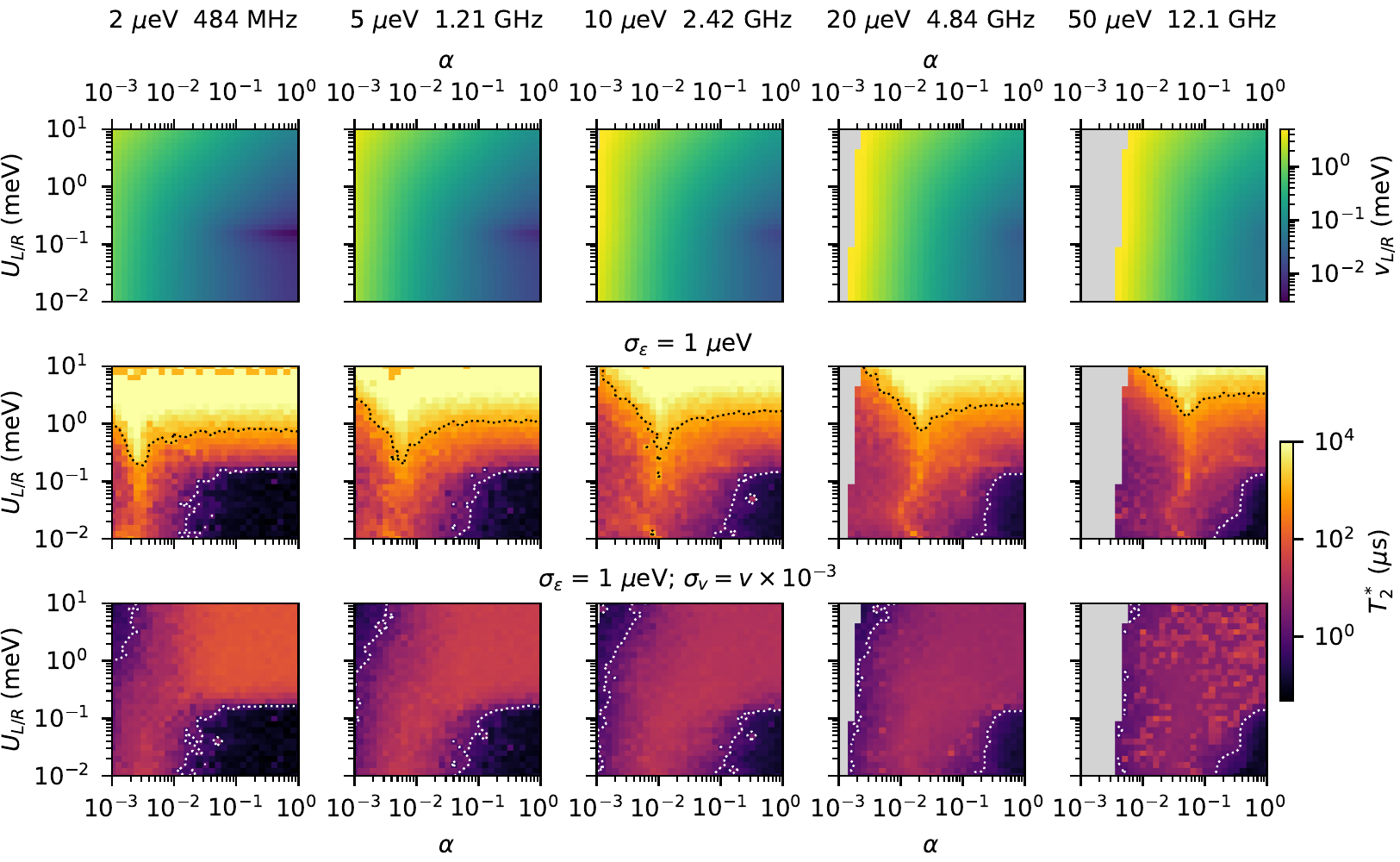}
	\caption{Coupling between the impurities required to achieve the splitting between the qubit states, and the estimated inhomogenous dephasing times for the two nose models. The columns 1 and 4 present the same parameter values as Fig.~7, albeit for fewer realizations of the noise.}
	\label{fig_tunability_overview}
\end{figure*}

\begin{itemize}
	\item (Fig.~\ref{fig_no_tunneling_overview}) Charge stability diagrams without dot-island coupling (c.f.~Fig.~\ref{fig_no_tunneling})
	\item (Fig.~\ref{fig_symmetric_tunneling_overview}) Charge stability with $\alpha=0$, for symmetrically varied $v_{L/R}$ (c.f.~Fig.~\ref{fig_symmetric_tunneling})
	\item (Fig.~\ref{fig_asymmetric_tunneling_overview}) Charge stability with $\alpha=0$, for asymmetrically varied $v_{L/R}$
	\item (Fig.~\ref{fig_tunability_overview}) Coupling terms required to tune the DIQ to a desired frequency, for a larger set of qubit frequencies, as well as the corresponding $T_2^*$ for chemical-potential-only noise, as well for a combination of chemical-potential and coupling noise. $T_2^*$ vas estimated based on 100 qubit samples per pixel, compared to 500 samples in the main text, resulting in increased sampling noise. (c.f.~Fig.~\ref{fig_tunability})
\end{itemize}

\bibliography{biblio}